# LOGISTICS AND TRADE FLOWS IN SELECTED ECOWAS COUNTRIES: AN EMPIRICAL VERIFICATION

ERIAMIATOE EFOSA FESTUS

DEPARTMENT OF ECONOMICS

EMAIL: eriamiatoefestus@gmail.com, feriamiatoe@yahoo.com

+1-773-704-8936

**ABSTRACT**

*This study investigates the role of logistics and its six components on trade flows in selected Economic Community of West Africa States (ECOWAS) countries. The impact of other macro-economic variables on trade flows was also investigated. Ten countries were selected in eight years period. We decomposed trade flows into import and export trade. The World Bank Logistics performance index was used as a measure of logistics performance. The LPI has six components, and the impact of these components on trade flows were also examined. The fixed-effect model was used to explain the cross-country result that was obtained. The results showed that logistics has no significant impact on both Import and export, thus logistics play no role on trade flows among the selected ECOWAS countries. The components of logistics except Timeliness of shipments in reaching the final destination ( CRC ), have no impact on trade flows. Income was found to be positively related to imports. Exchange rate, consumption and money supply, reserve and tariff have no significant impact on imports. Relative import price has an inverse and significant relationship with imports. GDP has a positive and significant impact on export trade. The study also found FDI, savings, exchange rate and labour to have insignificant impact on exports. Finally, we found that logistics is not a driver of trade among the selected ECOWAS countries. The study recommended the introduction of the single window system and improvement in border management in order to reduce the cost associated with Logistics and thereby enhance trade.*

## 1.0 INTRODUCTION

One area of trade that have received enormous attention within the field of international trade is that the cost of trade. this is often considered a critical factor because it's one of the factors that determines the worldwide competitiveness of a country's output. One factor that would affect the value of trade is logistics. Logistics commonly refers to organizing and coordinating the movements of material inputs, final goods, and their distribution. It was first used systematically for military purposes, but its use gradually spread to commercial endeavors, often referred to as logistics management. The Council of Supply Chain Management Professionals defines logistics management as that part of supply chain management that plans, implements, and controls the efficient, effective forward and reverse flow and storage of goods, services, and related information between the point of origin and the point of consumption to meet customers' requirements. In Africa, the value of doing business in the form of import and export has been on the high side (Portugal-Parez and Wilson, 2009). Generally, poor logistics performance can significantly hinder international trade and impact negatively the competitiveness of any country. Poor infrastructure, complex customs rules and regulations and lack of transparency between different public entities are likely to slow the movement of products within and across countries. To contribute to the literature on why developing countries have higher trade costs and on the average lag behind in global trade flows, this research assesses the impact of logistics on trade flows in ECOWAS countries. This research is motivated by the subsequent factors. One factor is the increasing importance of logistics, trade infrastructure and facilitation to trade costs and trade volumes in African countries and a second factor is the recent availability of knowledge on measures of logistics by the World Bank.

## 1.2 THE OBJECTIVE OF THE STUDY

This research investigates:

1. The role of trade logistics on international trade in Economic Community of West African States (ECOWAS) countries.
2. The impact of the six components of logistics on trade flows.
3. The extent to which poor-quality logistics constitute a barrier to trade in West Africa.
4. To investigate the impact of other macro-economic variables on flow of trade.
5. To suggest the way forward in logistics to enhance trade.

## 1.3 HYPOTHESIS OF STUDY

The hypothesis for this study is stated in terms of the null hypothesis. It is stated in terms of imports and exports.

➢ For import, it is stated as follows:

$H_o$: There is no significant relationship between logistics and imports.
$H_0$: There is no significant relationship between income and imports.
$H_0$: There is no significant relationship between exchange rate and imports.
$H_0$: There is no significant relationship between consumption and imports.
$H_0$: There is no significant relationship between money supply and import.
$H_0$: There is no significant relationship between tariff and imports.
$H_0$: There is no significant relationship between reserve and imports.
$H_0$: There is no significant relationship between price and imports.
$H_0$ : There is no significant relationship between the ability to tack and trace on imports

$H_0$ : There is no significant relationship between competence and quality of logistics on imports

$H_0$ : There is no significant relationship between the Ease and affordability of handling shipments on imports.

$H_0$ : There is no significant relationship between the Timeliness of shipments and imports.

$H_0$ : There is no significant relationship between the competence in the local logistics services on imports

$H_0$ : There is no significant relationship between Efficiency and effectiveness of processes by customs on imports

➢ For export, it is stated as:

$H_0$: There is no significant relationship between logistics and exports.
$H_0$: There is no significant relationship between FDI and exports.
$H_0$: There is no significant relationship between output and exports.
$H_0$: There is no significant relationship between savings and exports.
$H_0$: There is no significant relationship between exchange rate and exports.
$H_0$: There is no significant relationship between labor force and exports.

$H_0$ : There is no significant relationship between the ability to tack and trace on exports.

$H_0$ : There is no significant relationship between competence and quality of logistics on exports

$H_0$ : There is no significant relationship between the Ease and affordability of handling shipments on exports.

$H_0$ : There is no significant relationship between the Timeliness of shipments and exports.

$H_0$ : There is no significant relationship between the competence in the local logistics services on exports.

$H_0$ : There is no significant relationship between Efficiency and effectiveness of processes by customs on exports.

## 1.4   SIGNIFICANCE OF THE STUDY

This study tends to verify the extent to which logistics has been able to facilitate the flow of trade within ECOWAS countries. However, there have been several studies that have examined the role of logistics on trade. This empirical study is intended to contribute and clarify the lacks and expand the body of research already available on the relationship between logistics and international trade.

## 1.5 Scope of The Study

This study attempts to examine the role of Logistics on the flow of trade in ECOWAS countries; ten countries were selected in this region. The scope of the study spares through 2007-2014, secondary data collected from the World Bank was used in this study. Fixed effect and random effect and the pooled OLS were used in the analyses.

## 2.0 Conceptual Review

### 2.1 Logistics Performance Index (LPI)

To capture the rate of performance of logistics of different countries, the World Bank developed the logistics performance index in 2007. The index is measured on a scale of 1 to 5. Scale 1 and 5 represent low and high performance, respectively. The index is the average of scores covering six sub-dimensions (Arvin et al 2014). According to World Bank the six sub-dimensions to benchmark countries logistics performance are:

1. Efficiency and effectiveness of processes by customs and border agencies at the borders.

2. Quality of transport-related and IT infrastructure.

3. Ease and affordability of handling shipments in and outside the country.

4. Competence in the local logistics services industry.

5. Ability to track and trace shipments throughout the logistics chain.

6. Timeliness of shipments in reaching the final destination.

Logistics performance in Africa has been reportedly poor, which may be attributable to the poor state of infrastructure, weak institutions, technological deficiency and administrative bottlenecks. This has led to high cost of goods and low competitiveness of goods produced in Africa in the foreign markets.

To support this position, the logistics performance index for in some Africa in Countries in 2010 is presented in the table below:

**Table 2.1**

| COUNTRY | SCORE | CUSTOM | INFRASTRUCTURE | INTERNATIONAL | LOGISTICS | TRACKING | TIMELINESS |
|---------|-------|--------|----------------|---------------|-----------|----------|------------|

|  |  |  |  | SHIPMENT | COMPETENCE | AND TRACING |  |
|---|---|---|---|---|---|---|---|
| South Africa | 3.4 | 3.22 | 3.42 | 3.26 | 3.59 | 3.73 | 3.57 |
| Senegal | 2.86 | 2.45 | 2.64 | 2.75 | 2.73 | 3.08 | 3.52 |
| Uganda | 2.82 | 2.84 | 2.35 | 3.02 | 2.59 | 2.45 | 3.52 |
| Tanzania | 2.60 | 2.42 | 2.00 | 2.78 | 2.38 | 2.56 | 3.33 |
| Kenya | 2.59 | 2.23 | 2.14 | 2.84 | 2.28 | 2.89 | 3.06 |
| Nigeria | 2.59 | 2.17 | 2.43 | 2.84 | 2.45 | 2.45 | 3.10 |
| Cameroon | 2.55 | 2.11 | 2.10 | 2.69 | 2.53 | 2.60 | 3.16 |
| Cote d'Ivoire | 2.53 | 2.16 | 2.37 | 2.44 | 2.57 | 2.95 | 2.73 |
| Ghana | 2.47 | 2.35 | 2.52 | 2.38 | 2.42 | 2.51 | 2.67 |
| Ethiopia | 2.41 | 2.13 | 1.77 | 2.76 | 2.14 | 2.89 | 2.65 |
| Zambia | 2.28 | 2.17 | 1.83 | 2.41 | 2.01 | 2.35 | 2.85 |
| Angola | 2.25 | 1.75 | 1.69 | 2.38 | 2.01 | 2.54 | 3.01 |
| SSA Average | 2.42 | 2.18 | 2.05 | 2.51 | 2.28 | 2.49 | 2.94 |

Source: World Bank IPI Index 2010

These indicators are representative of the views of a large range of logistics providers and logistics buyers. Selection of indicators was based on interviews with professionals in international freight logistics. The data was gathered from managerial level personnel of international freight forwarding firms worldwide. The perceptions are therefore representative of the views of a large range of logistics providers and logistics buyers

In 2014, Nigeria and Côte d'Ivoire were the strongest trade facilitation performers in West Africa. However, their overall LPI scores are still significantly lower than that of South Africa. By contrast, The Gambia, Niger, and Togo are the weakest performers, with scores well below the West African average of 2.54 (compared with 3.43 for South Africa). Trade facilitation is therefore a very real and serious constraint on trade, both intraregional and extra-regional.

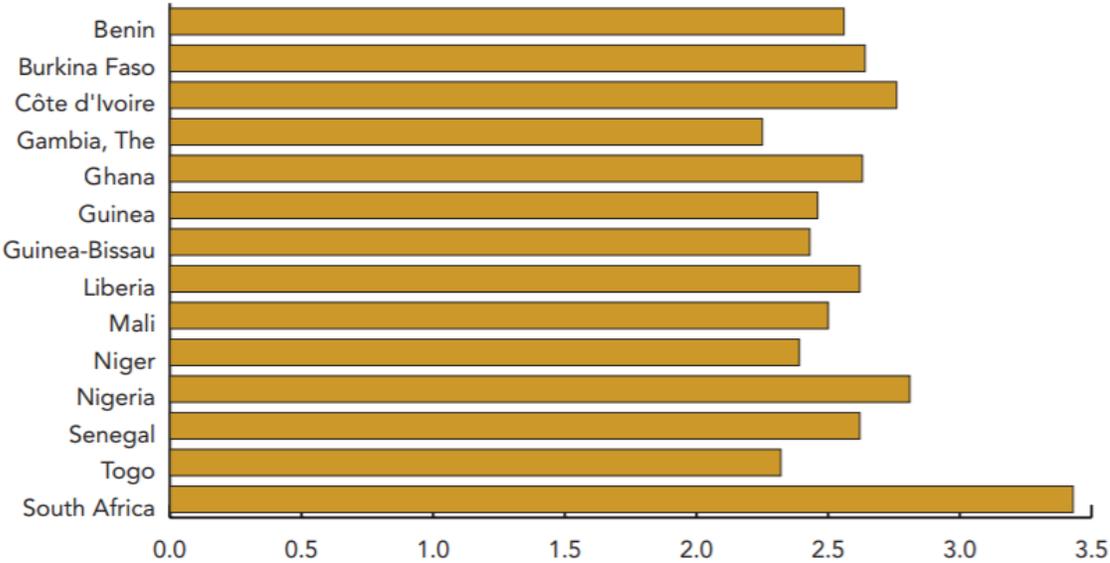

Figure 2.1 Overall Logistics Performance Index Scores for ECOWAS Member States and South Africa, 2014

Source: Logistics Performance Index, World Bank database 2014.

Figure 2.2 presents ECOWAS averages for each component of the LPI, again compared with South Africa as a performance benchmark. West Africa clearly lags regional best practice in every area of logistics performance. The gap is largest in relative terms in logistics services and infrastructure. That result means that the quantity and quality of infrastructure in West Africa are holding back trade, and the same is true of private logistics services markets, including transport and freight forwarding. The policy agenda that those results suggest therefore needs to include both public sector investment (in infrastructure development and maintenance) and private sector development (of relevant services sectors). Of course, private sector development itself requires an appropriate regulatory stance that encourages market-based competition. Openness to trade in services and foreign direct investment can be one part of an overall policy mix that encourages the development of logistics and trade facilitation.

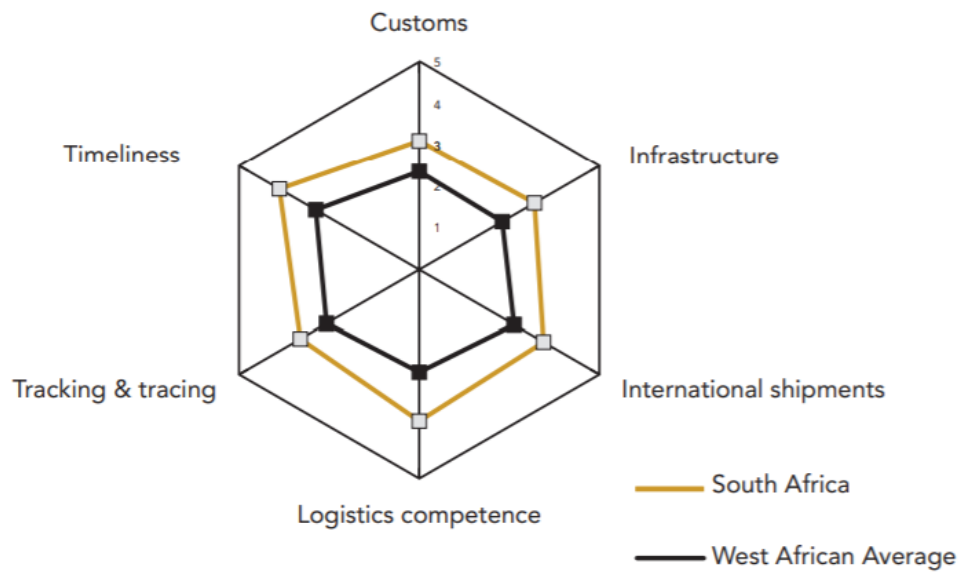

Figure 2.2 Logistics Performance Index Component Scores for ECOWAS Member States (simple average) and South Africa, 2014

Source: Logistics Performance Index, World Bank database 2014.

## 2.2 Empirical Review

There have been several researches carried out to determine the role of logistics on trade flows in different part of the world.

Julia Devlin and Peter Yee (2002), using a mathematical model, investigates the impact on foreign trade from industry logistics system and the results showed that the logistics efficiency would affect the time cost of trade, furthermore, affect the trade effects.

Michael P. Keane and Susan E. Feinberg (2007) took Canada and United States as study sample, and found that the trade costs decline were conducive to the trade exchanges between enterprises. Among the decline of trade costs, logistics costs' decline played an important role.

Among the infrastructure necessary to facilitate trade, the efficiency of ports has received specific attention in Sanchez et al (2003) and Clark et al (2004). These authors show that port efficiency is relevant for a large proportion of transactions related to international trade. This is true not only for the activities that depend directly on port infrastructure, such as pilotage, towing, stevedoring, or even freight storage and depositing, but also for other administrative activities, including fulfilling customs requirements.

Luttermann and Kotzab (2017) examined the impact of logistics on international trade and investment flows. The scholars used a panel data containing 20 Asian countries over the years 2006 to 2014. Logistics performance index and the global competitiveness index were used to illustrate the logistic system of countries. They found a statistically significant relationship between logistics and trade. They, however, found logistics do not seem to contribute to the ability of the countries to attract FDI.

Sanchez *et al* (2003) measure port efficiency using data on efficiency in time, port productivity and vessel length of stay at port obtained from surveys sent to 55 Latin American port terminals in 1999. The results obtained indicate that an increase in port productivity reduces transport costs.

Azat Gani (2017) carried out an investigation of the effect of logistics performance in international trade. The empirical analyses adopts the cross-sectional estimation phase combine with cross-country and time-series data involving sixty countries and four-time period. The findings reveal that the overall logistics performance is positively related and significantly correlated with exports and imports.

Nunez-Sanchez and Coto-Millan (2012) calculate an index of technical efficiency for Spanish ports and prove that despite this index averaging 78.6% for the port system as a whole (for the period 1986-2005), there are considerable differences between ports with those in Valencia, Tenerife and Algeciras being the most efficient. This study uses data gathered from the Annual Reports of Puertos del Estado (several years).

According to Arvis *et al*, (2007), the LPI suggests that there are strong synergies among reforms to customs, border management, infrastructure, and transport regulations. Thus strong logistics performance is associated with increased trade in developing countries.

Oualid Kherbash and Marian Liviu (2015) reviewed and identified the essential factors that affect logistics and transport sector through globalization process happening in the world economy. They conclude that transport substantially supports international economic relations and plays a primary part in creating a world network of exchange of goods and in the transfer of capital goods among countries in the middle of several transport modes.

Wilson *et al* (2004) quantified and examined the impact of trade facilitation on trade costs and volumes. The authors find unilateral trade facilitation reforms in the areas of port efficiency, customs and regulatory environment reforms and e-business to be significant determinants of increasing trade flows.

Wei Ma, Xiaoshu Cao and Jiyuan Li (2021) examined the impact of logistics development level on international trade in China. They analyzed data from 31 China Provinces to 65 Countries by using the improved gravity model from 2008 to 2018. The result revealed that logistics development level had significantly promoted international trade development.

Muhammad M. Hamed (2019) explored the performance of the logistics sector in Jordan using the logistics performance index of six dimensions over eight years life span. The findings reveal that the Jordan LPI has dropped significantly in 2018 when compared to 2016.

Using a measure of port efficiency provided by the 1999 Global Competitiveness Report of the World Economic Forum, the results obtained by Clark *et al* (2004) indicate that this variable has a marked impact on international trade through transport costs. More specifically, they find that an improvement in port efficiency from the 25th percentile to the 75th percentile reduces maritime transport costs by more than 12%, equivalent to 5,000 miles in terms of geographical distance, and would entail a 25% rise in bilateral trade.

Wang and Liu (2014) using the error correction model, studied the relationship between port logistics development and agricultural trade. The result revealed that Port Logistics development have a positive role in promoting the development of agricultural product trade both in the short and long term.

Dollar and Kraay (2002 & 2004), Rodrik et al (2004) have provided evidence to the effect that institutions as well as infrastructure and facilitation matters for trade and that if some countries were lagging behind in terms of trade and growth it had something to do with the poor state of institutions and infrastructure among other factors. This consensus has informed the development agenda of Development Agencies in the developing world who have in recent times focused on trade facilitation and institutional building to improve trade.

**3.0        Theoretical Framework**
Basically, this research was built on the Gravity model theory.

**3.1    Gravity Model Theory**

It has been known since the seminal work of Tinbergen (1962) that the dimensions of bilateral trade flows between any two countries are often approximated by a law called the "gravity equation" by analogy with the Newtonian theory of gravitation even as planets are mutually attracted in proportion to their sizes and proximity, countries trade proportion to their respective GDPs and proximity. Initially, the gravity equation was thought of merely as a representation of an empirically stable relationship between the dimensions of economies, their distance, and therefore the amount of their trade. The extraordinary stability of the gravity equation and its power to elucidate bilateral trade flows prompted the look for a theoretical explanation for it. Whereas empirical analysis predated theory, we all know now that the majority trade models require gravity so as to figure. the primary important plan to provide a theoretical basis for gravity models was the work of Anderson (1979). He did so within the context of a model where goods were differentiated by country of origin and where consumers have preferences defined over all the differentiated products. This structure would imply that, regardless of the price, a Country will consume a minimum of a number of very good from every country. All goods are traded, all countries trade and, in equilibrium, value is that the sum of home and foreign demand for the unique good that every country produces. For this reason, larger countries import and export more. Trade costs are modelled as "iceberg" costs, that is, only a fraction of the freight arrives to destination, the remainder having melted in transit. Clearly, if imports are measured at the CIF value, transport costs reduce trade flows. Deardorff (1998) shows that a gravity model can arise from a standard factor proportions explanation of trade. Eaton and Kortum (2002) derive a gravity-type equation from a Ricardian sort of model, and Helpman et al. (2008) and Chaney (2008) obtained it from a theoretical model of international trade differentiated goods with firm heterogeneity. In its general formulation, the gravity equation is given as :

$$F_{ij} = G(M_i^{\beta_1} M_j^{\beta_2} / D_{ij}^{\beta_3}) \quad \quad 3.1$$

Where,

F is the trade flows.

M is the economic mass of each country.

D is the distance.

G is the constant.

The contribution of recent research concerning the theoretical foundation of the gravity equation is to possess highlighted the importance of deriving the specifications and variables utilized in the gravity model from theory so as to draw the right inferences from estimations of gravity equation. However, Anderson and Van Wincoop (2003), argued that by not taking under consideration multilateral resistance terms (i.e., relative prices), the normal gravity equation had not been correctly specified. The motivation behind this argument stemmed from the highly overstated impact of national borders found by McCallum (1995) resulting from estimating the normal gravity equation for bilateral trade between us and Canada. McCallum (1995) estimated an equation model for U.S. states and provinces of Canada with two z variables (bilateral distance and a dummy variable that is equal to one of the 2 regions are located within the same country and equal to zero otherwise). After controlling for distance and size, McCallum found trade between provinces to be twenty-two times greater than trade between states and provinces, suggesting that there have been substantial trade costs incurred in trade across the United States-Canada border. Anderson and van Wincoop (2003) theory-based gravity equation was, therefore, a theoretical refinement of the normal gravity model to incorporate multilateral trade resistance variables. As suggested by Anderson and Van Wincoop (2003) and Feenstra (2004), a method of augmenting the normal gravity equation with multilateral resistance terms is to incorporate exporter and importer fixed effects resulting in the stochastic theory-based gravity equation of the form;

$$X_{ij} = \emptyset_0 \, Y_j^{\varphi 1} \cdot Y_j^{\varphi 2} \cdot Z_{ij}^{\varphi 3} \cdot e^{\alpha \, 1di + \alpha 2dj} \quad \quad 3.2$$

Equation 3.2 can be reformed as :

$$X_{ij} = \emptyset_0 Y_j^{\varphi_1} \cdot Y_j^{\varphi_2} \cdot L_{ij}^{\varphi_3} \cdot e^{\alpha_1 d_i + \alpha_2 d_j} \text{ -----------------------------------------------------------------------3.3}$$

Where $\varphi_0, \varphi_1, \varphi_2, \varphi_3, \alpha_1$ and $\alpha_2$ are unknown parameters to be estimated, and $d_i$ and $_{DJ}$ are exporter and importer dummies and $\varphi_1 = \varphi_2 = 1$ (unit-income elastic). $X_{ij}$ is the volume of trade, $Y_{ij}$ is the economic mass (GDP), $Z_{ij}$ is the distance and $L_{ij}$ is Logistics of both the exporting the importing countries. The Anderson and van Wincoop (2003) theory-based gravity equation has been mainly employed by various authors to elucidate the pattern of bilateral trade amongst countries. In addition to augmenting the normal gravity equation with multilateral resistance terms in an effort to completely explain bilateral trade amongst countries, the normal specification also because the theory-based gravity equation has been subjected to further augmentation to incorporate other factors that are deemed significant determinants of trade costs and volumes. Most studies that have made use of the gravity equation have augmented it with various measures of distance and country characteristics, also as measures of trade facilitation, infrastructure, and logistics. Conclusively, distance within the gravity model is inversely associated with trade flows. Trade cost (cost of transportation, cost of clearance, warehousing, cost of tracing and tracking etc) is additionally inversely associated with trade flows. Hence, cost of transportation, cost of clearance, warehousing, tracing and tracking are associated with logistics.

### 3.2    Model Specification

Drawing the theoretical framework and from the host of factors that has been identified as determinants of international trade from the several empirical and theoretical literatures reviewed. Trade flows was decomposed into export trade and import trade. The import and export trade formed the dependent variables of the four separate models. The impact of the six components of logistics on both export and imports were examined.

Model for Import trade can be expressed as:

IMPT= f (LOG, GDP, EXCH, CONS, MS, TARF, RES, PRICE) ------------------------------------3.4

Where,

IMPT: Import trade
LOG: Logistics
GDP: Gross Domestic Product
EXCH: Exchange rate
CONS: Consumption
MS: Money supply
TARF: Tariff rate
RES: Foreign reserve
PRICE: Relative price
$\alpha_i$ = Country specific time-invariant factor
$\gamma_t$ = Year-specific fixed effects

Equation 3.3 can be expressed in econometric (log) form as:

$$LNIMPT_{it}=\alpha_0+\gamma_t+\alpha_1 LNLOG_{it}+\alpha_2 LNGDP_{it}+\alpha_3 LNEXCH_{it}+\alpha_4 LNCONS_{it}+\alpha_5 LNMS_{it}+\alpha_6 LNTARF_{it}+\alpha_8 LNRES_i+\alpha_9 LNPRICE_{it}+\mu_{it} \quad \text{------3.5}$$

Where the subscript i (=1…n) represent the countries and t (=1…i) represents the period of time (years).

The aprior expectations are as follows:

$\alpha_1, \alpha_2, \alpha_4, \alpha_5, \alpha_7 > 0 \quad \alpha_3, \alpha_6, \alpha_8 < 0$

The model for Export can be expressed as:

EXPT= f (LOG, FDI, GDP, SAV, EXCH, LF,) ------3.6

Where,
EXPT: Export trade
LOG: Logistics
FDI: Foreign Direct Investment
GDP: Gross Domestic Product
SAV: Domestic savings
EXCH: Exchange rate
LF: Labour force
$\beta_i$ = Country specific time-invariant factor
$\gamma_t$ = Year-specific fixed effects

Equation 3.5 can be expressed in econometric (Log) form as:

$$LNEXPT_{it} = \beta_0+\gamma+\beta_1 LNLOG_{it}+\beta_2 LNFDI_{it}+\beta_3 LNGDP_{it}+\beta_4 LNSAV_{it}+\beta_5 LNEXCH_{it}+\beta_6 LNLF_{it}+\mu_{it} \quad \text{------3.7}$$

Where the subscript i (=1…n) represent the countries and t (=1…i) represents the period of time (years).

The aprori expectation are as follows:

$\beta_1, \beta_2, \beta_3, \beta_4, \beta_6, > 0 \quad \beta_5 < 0$

IMPT= f (TNT, QLS, CPS, ECC, CRC, QTT) ------3.8

These can be expressed in econometric (log) form with respect to imports and export as :

$$LNIMPT_{it}=\delta_0+\gamma_t+\delta_1 LNTNT_{it}+\delta_2 LNQLS_{it}+\delta_3 LNCPS_{it}+\delta_4 LNECC_{it}+\delta_5 CRC_{it}++\delta_6 QTT_{it}+\mu_{it} \quad \text{-----3.9}$$

EXPT= f (TNT, QLS, CPS, ECC, CRC, QTT) ------3.10

$$LNEXPT_{it}=\pi_0+\gamma_t+\pi_1 LNTNT_{it}+\pi_2 LNQLS_{it}+\pi_3 LNCPS_{it}+\pi_4 LNECC_{it}+\pi_5 CRC_{it}++\pi_6 QTT_{it}+\mu_{it} \text{--3.11}$$

Where,
IMPT: Imports
EXPT: Export
TNT: Ability to track and trace shipments throughout the logistics chain
QLS: Competence in the local logistics services industry
CPS: Ease and affordability of handling shipments in and outside the country
ECC: Efficiency and effectiveness of processes by customs and border agencies at the borders

CRC: Timeliness of shipments in reaching the final destination.
QTT: Quality of transport-related and IT infrastructure.
$\alpha_i$ = Country specific time-invariant factor
$\gamma_t$ = Year-specific fixed effects
Where the subscript i (=1…n) represent the countries and t (=1…i) represents the period of time (years).

The aprori expectation are as follows:
$\delta_1, \delta_2, \delta_3, \delta_4, \delta_5, \delta_6, > 0$

The aprori expectation are as follows:
$\pi_1, \pi_2, \pi_3, \pi_4, \pi_5, \pi_6, > 0$

### 3.3 Research Design And Methodology

The research design adopted for this study is a cross-country research design. Ten countries in ECOWAS were selected; Nigeria, Ghana, The Gambia, Sierra-Leone, Senegal, Niger, Mauritania, Benin, Burkina-Faso and Guinea. The research work employed basically the secondary data sourced from World Bank. The period of estimation is 2007-2014. The data relating to currency was expressed in current U.S Dollars.

In this study, the method of data analysis is the Fixed Effects or Random Effects analysis. The Hausman test will be used to determine which model will be adopted in our analysis – Fixed effect or Random effect models.

### 4.1 Presentation of Empirical model

This study focuses on selected ECOWAS countries. We utilized the cross-section of ten ECOWAS countries for the period of 2007-2014. Therefore, a total of eight observations were utilized. The study utilized a combination of a panel least square, fixed effect and random effect model. In addition, the Hausman test was utilized as the heuristic device that helps us in making choice between the fixed effect and the random effect model (Iyoha, 2014; Urhoghide and Emeni, 2014).

In Table 4.1 below, we present the estimated models using the three statistical approaches.

Table 4.1: Panel Regression result for Imports

| Variables | Pooled OLS | Fixed Effect | Random Effect |
|---|---|---|---|
| LNLOGS | 0.085481 | 0.080059 | 0.292774 |
|  | (0.5557) | (0.6157) | (0.2691) |
| LNGDP | 0.965406 | 0.905630 | 0.929335 |
|  | (0.0009) | (0.0067) | (0.0081) |
| LNEXCH | -0.017174 | 0.196636 | -0.109066 |
|  | (0.8660) | (0.5448) | (0.1770) |
| LNCONS | -0.194328 | -0.055646 | -0.185193 |
|  | (0.4735) | (0.8575) | (0.5896) |
| LNMS | -0.019808 | 0.003233 | 0.115568 |
|  | (0.7386) | (0.9583) | (0.1263) |
| LNTARF | 0.765880 | 0.553848 | 0.294526 |
|  | (0.3201) | (0.4874) | (0.5882) |
| LNRES | -0.072491 | -0.084678 | 0.070298 |
|  | (0.0866) | (0.0754) | (0.0917) |
| ENRIC | -0.428564 | -0.392020 | -0.140002 |
|  | (0.0000) | (0.0007) | (0.2201) |
| C | 6.635203 |  | 0.462793 |
|  | (0.0876) |  | (0.8372) |
|  | (0.0000) | (0.0000) |  |
| R-squared | 0.991930 | 0.993107 | 0.852358 |
| F-Statistic | 805.8183 | 400.2298 | 50.51480 |

Note: (    ) represents the probability value of the coefficients

Source: Author's computation

The table above shows the result for the model which examine the impact of certain hypothesized variables on imports for selected ECOWAS countries.

In order to examine the impact of the regressors on the dependent variable, we examine the estimated coefficients. Logistics (LNLOGS) has 0.085481, 0.080059 and 0.292774 coefficients in the three estimated models, with probability values of 0.5557, 0.6157 and 0.2691 respectively in the pooled OLS, fixed effect and the random effect models. Hence Logistics is statistically insignificant in the estimated models, given that their respective probability values are less than 0.05.

Income proxied by GDP has a positive impact on imports in the three estimated models and the estimated coefficients are respectively 0.965406, 0.905630 and 0.929335 in the pooled OLS, fixed effect and the random effect models. And the probability values of income in the three models are less than 0.05, hence income is statistically significant at 5% level. The coefficient for exchange rate are respectively -0.017174, 0.196636 and –0.109066 in the pooled OLS, fixed effect and the random effect models, respectively. The probability values for the exchange rate for the three estimated coefficient exceed 0.05, hence exchange rate is statistically insignificant at 5% level. The coefficients of consumption in the three models are given respectively as -0.194328, -0.055646 and -0.185193 in the pooled OLS, fixed effect and random effect models. The probability values of the three estimated models exceed 0.05, consumption is thus statistically insignificant in the three models. The coefficients for money supply in the pooled OLS, fixed effect and the random effect models are -0.019808, 0.003233 and 0.115568 respectively. The probability values of the three estimated models exceed 0.05, hence Money supply is statistically insignificant at 5% level. The coefficient of tariff of the pooled OLS, fixed effect and the random effect are 0.765880, 0.553848 and 0.294526 respectively. The probability values for the three estimated values exceed 0.05, hence tariff is statistically insignificant in the three estimated models. The coefficient of reserve in the pooled OLS, fixed effect and random effect are -0.072491, -0.084678 and 0.070298, respectively. And the probability values for the three estimated models exceed 0.05, hence reserve is statistically insignificant in the three estimated models. The coefficient of price in the -0.428564, -0.392020 and -0.040002 in the pooled OLS, fixed effect and random effect models. The probability values of the pooled OLS and fixed effect models are less than 0.05, hence price is statistically significant in both models at 5% level but greater 0.05 in the random effect, hence statistically insignificant in the random effect model. However, if we go by the identification test that is the Hausman's chi-square statistic (2=36.94) with probability value of 0.0000, then the fixed effect model is more reliable hence shows higher explanatory power.

Table 4.2: Panel regression result for Export

| Variables | Pooled OLS | Fixed Effect | Random Effect |
| --- | --- | --- | --- |
| LN(LOGS) | 0.003743 | 0.099065 | 0.022667 |
| | (0.9885) | (0.7111) | (0.9587) |
| LN(FDI) | -0.039195 | -0.048352 | 0.130236 |
| | (0.3636) | (0.3323) | (0.0061) |
| LN(GDP) | 0.943471 | 0.901530 | 1.187617 |
| | (0.0000) | (0.0006) | (0.0000) |
| LN(SAV) | 0.062517 | 0.080889 | 0.051410 |
| | (0.3441) | (0.2433) | (0.5433) |
| LN(EXCH) | -0.063873 | -0.107063 | 0.011691 |
| | (0.4128) | (0.7999) | (0.7185) |
| LN(LF) | -0.301838 | 2.339975 | -0.474444 |
| | (0.2807) | (0.2023) | (0.0048) |
| C | 4.755188 | | -2.142972 |
| | (0.1265) | | (0.0471) |
| $R^2$ | 0.9911271 | 0.992436 | 0.894909 |
| F-Statistic | 827.3278 | 376.1355 | 90.83309 |

Note: ( ) represents the probability values for the coefficient

Source: Author's Computation

Examining the impact of the individual regressors on the dependent variables, we examine the estimated coefficients. As can be observed, Logistics coefficients for the three estimated models are 0.003743, 0.099065 and 0.022667 respectively. The probability values for the three estimated models exceed 0.05, hence Logistics is statistically insignificant in the three estimated models. The coefficients of of FDI in the pooled OLS, fixed effect and the random effect models are -0.039195, -0.048352 and 0.130236. And the probability values of the pooled OLS and fixed effect models exceed 0.05 but less than 0.05 in the random effect model. Thus FDI is statistically insignificant in the pooled OLS and the fixed effect models but statistically significant in the random effect models. The coefficient of GDP in the pooled OLS, fixed effect and the random effect are 0.943471, 0.901530 and 1.187617 respectively. The probability values for the three estimated models are 0.0000, 0.0006 and 0.0000 respectively.

Thus GDP is statistically significant at 5% in the three estimated models. Thus GDP is a critical driver of export trade in the selected ECOWAS countries. The coefficients of savings for the pooled OLS, fixed effect and the random effect are 0.062517, 0.080889 and 0.051410 respectively, they are correctly signed. The probability values of the three estimated models are greater than 0.05, hence savings is statistically insignificant at 5% levels in the models estimated. The coefficient of exchange rate for the three estimated models are -0.063873, -0.107063 and 0.011691. The probability values for the three estimated models exceed 0.05, hence exchange rate is statistically insignificant at 5% levels. The coefficient for labour force in the three estimated models are -0.301838, 2.339975 and -0.474444 respectively. The probability value for the pooled OLS and the fixed effect exceed 0.05, hence statistically insignificant at 5% levels but less than 0.05 in the random effect model, hence statistically significant in the random effect model at 5% levels.

Furthermore, if we go by identification test that is Hausman's chi-square statistic (2 =20.024672, p=0.0027), then the fixed effect model is more reliable and performed better as reflected by the coefficient of determination. This suggest that the causal relationship in our model is influenced by a cross-section specific effects which are realization of independent random variables with mean zero and finite variance and uncorrelated with idiosyncratic residual (Ayemere and Afensimi, 2014)

Table 4.3: Panel Regression result for Imports

| Variables | Pooled OLS | Fixed Effect | Random Effect |
| LN(TNT) | 0.21810 | 1.26263 | 0.21810 |
| | (0.9377) | (0.67835) | (0.93720) |
| LN(QLS) | -3.95615 | -4.96762 | -3.95615 |
| | (0.1459) | (0.09892) | (0.1358) |
| LN(CPS) | 2.50543 | 2.86514 | 2.50543 |
| | (0.2874) | (0.24728) | (0.2791) |
| LN(ECC) | 0.55839 | 0.74437 | 0.55839 |
| | (0.8221) | (0.77616) | (0.8207) |
| LN(CRC) | 4.67591 | 4.72039 | 4.67591 |
| | (0.0538) | (0.06182) | (0.0450) |
| LN(QTT) | 1.87623 | 1.39691 | 1.87623 |
| | (0.3911) | (0.56051) | (0.3844) |
| C | 16.16714 | | 16.16714 |
| | (0.6644) | | (0.1933) |
| | 0.25499 | 0.25499 | 0.23928 |

Note: (    ) represents the probability values for the coefficient

Source: Author's Computation

From the result above, if we go by using the Hausman test with chi-square value of 1.0484 and with probability value of 0.9837 which shows that the Random effect model is better. Examining the regressors , We found that all the variables are not significant except CRC because their probabilities are greater than 0.05. CRC has a coefficient of 4.67591with probability value of 0.0450.

Table 4.4: Panel Regression result for Export

| Variables | Pooled OLS | Fixed Effect | Random Effect |
|---|---|---|---|
| LN(TNT) | -0.46675 | 0.43358 | -0.46675 |
|  | (0.88029) | (0.89807) | (0.87931) |
| LN(QLS) | -2.22032 | -2.74969 | -2.22032 |
|  | (0.45678) | (0.40329) | (0.45110) |
| LN(CPS) | 2.76816 | 2.91732 | 2.76816 |
|  | (0.28996) | (028904) | (0.28165) |
| LN(ECC) | -1.07221 | -1.29393 | -1.07221 |
|  | (0.69782) | (0.65742) | (0.69514) |
| LN(CRC) | 4.74205 | 4.74075 | 4.74205 |
|  | (0.07685) | (0.09019) | (0.06722) |
| LN(QTT) | 2.04052 | 1.29598 | 2.04052 |
|  | (0.0094) | (0.62731) | (0.39441) |
| C | 15.77375 |  | 15.77375 |
|  | (0.07301) |  | (0.06296) |
| R-Squared | 0.17814 | 0.091094 | 0.17814 |

Note: (    ) represents the probability values for the coefficient

Source: Author's Computation

From the result above, if we go by using the Hausman test with chi-square value of 1.0484 and with probability value of 0.9837 which shows that the Random effect model is better. Examining the regressors, we found that all the explanatory variables have probability values greater than 0.05. Hence the explanatory variables are not statistically significant.

In summary, given that the components of logistics are statistically insignificant could be the reason why Logistics performance index (LPI) is not statistically significant in explaining trade flows (export and imports).

## 4.2    POLICY IMPLICATIONS

Based on the empirical estimation, it has been observed that Logistics is statistically insignificant in both imports and exports models. This depicts that Logistics has no significant impact on both import and export trade in the selected ECOWAS countries under study. This contradicts the findings in many empirical studies in literature, however from other parts of the world. Like Kun (2001), Clark et al (2004), Wilson et al (2005), among other studies. This may be attributable to the fact that ECOWAS are developing hence faced with the challenge of poor infrastructures, administrative bottleneck, technological deficiency, and weak institution. Dollar Cray (2002) Rodrik et al (2014) and Chang et al (2005) have provided evidence that institutions, infrastructures and facilitation matter in trade and stated that if some countries were lagging behind in terms of trade and growth, it has something to do with poor state of infrastructure and institutions among others.

According to World Bank (2007), most countries in need of attention from the international community and their neighbors are those with government challenges such as post-conflict states and fragile states as well as those challenged in their economic or geography in their connectivity to global markets such as landlocked developing countries and small highland states.

Income proxy by GDP was found to have a positive relationship with imports. The implication is that a percentage increase in income will lead to about 9.6% increase in imports. This shows that an increase in income encourages importation. For the export model, Output was proxied by GDP. The result shows that output is positively and significantly related to export trade in the selected ECOWAS countries. The result shows that a percentage increase in output will result to 9.01% increase in exports. Exchange rate is statistically insignificant in both import and export models. This depicts that exchange rate plays no role on trade flows in the selected ECOWAS countries. Consumption was found to be statistically insignificant, hence has no significant impact on imports. Money supply, tariff, and reserve were found to be statistically insignificant, hence have no significant impact of trade flows in the selected ECOWAS countries. This may be attributable to the low level and continuous dwindling of reserves of most ECOWAS countries compared to huge reserves of developed nations of the world. Price was found to be negatively and significantly related to import trade. This is in conformity with the economic theory. Thus a percentage increase in price will result to -3.920% decrease in imports. FDI is statistically insignificant, hence has play no significant role on export trade in ECOWAS countries selected. This conforms to the arguments of the Antagonist of FDI. They argued that FDI is a clog in the wheel of development. They argued that the motives of foreign direct investors is resource soaking and repatriation of capital, thereby worsening the economic state of the host country rather than facilitate development. However, this contradicts the findings of Pfaffermayr (1996). Savings and Labour force are statistically insignificant, hence has not played any significant role on exports. This may be attributable to the fact that ECOWAS countries are characterized by low capital formation and poor labour, which result in poor productivity.

A percentage increase in Timeliness of shipments in reaching the final destination (CRC) will increase imports by 4.6 percent. The other five-component are not statistically significant, hence does not explain trade flows in the ECOWAS region.

## 4.3    SUMMARY OF THE FINDINGS

This study attempts to examine the role of logistics on trade flows in selected ECOWAS countries.

In the study, trade flow was decomposed into imports and exports trade and logistics was proxied by logistics performance index and the impact of other macro-economic variables on imports and exports were also examined.

The study revealed that logistics play no significant role on imports and exports in the selected ECOWAS countries. Among the components of logistics examined only the CRC is statistically significant with respect to import trade. Income has a positive and significant relationship with imports. Exchange rate, money supply, reserve and tariff have no significant relationship with imports. Price has an inverse and significant relationship with imports.

Income shows a positive and most significant relationship with imports hence an increase in income will lead to increase in imports and also Output proxied with GDP was found to be positively related to exports.  FDI, savings, exchange rate, savings and labour force have no significant impact on exports.

## 4.4    RECOMMENDATION

From the study, it has been observed that there is no significant relationship between Logistics and some other macro-economic variables on imports and exports. The following recommendation should be borne
1. Efficient border management is critical for eliminating avoidable delays and enhancing predictability in border clearance. Coordination among government control agencies will remain essential in trade facilitation efforts by introducing best practices in automation and risk management in custom and non-custom control agencies.
2. Transport infrastructures should be properly maintained to ensure easy carriage of goods from industries to the markets.
3. Stabilization of macro-economic variables that affects trade flow. The government through its monetary authorities pursue a stable exchange rate. The income level should be increased via an increase in productivity in order to enhance exports.

4. More labour should be employed and properly trained in order to increase productivity, thereby enhancing exports.
5. To improve exports, governments should seek for more foreign investors with the right economic motives.

**4.5 CONCLUSION**

In this study it has been observed that logistics plays no role on trade flow in the ECOWAS countries. Some macro-economic variables such as FDI, exchange rate, savings, consumption, money supply, reserve and tariff were not significant determinants of trade in ECOWAS countries. GDP was also found to be the major determinant of export trade in ECOWAS countries.

We can confidently say if the policy measures enumerated above are adhered to, trade-in ECOWAS countries will be improved, and cost reduced.